\documentclass[11pt,reqno,a4paper]{amsart}
\usepackage{amsmath}
\usepackage{amssymb}
\usepackage{amsfonts}
\usepackage[ansinew]{inputenc}
\usepackage{graphicx}
\usepackage{color}

\newtheorem{thm}{Theorem}

\newtheorem{lem}{Lemma}
\newtheorem{prop}{Proposition}
\theoremstyle{definition}

\newcommand{\be}{\begin{eqnarray}}
\newcommand{\ee}{\end{eqnarray}}
\newcommand{\dep}{\partial}
\def\eqq{\stackrel{\Sigma}{=}}
\def\neqq{\stackrel{\Sigma}{\neq}}

\begin{document}

\title{Cylindrically symmetric static solutions of the Einstein field equations for elastic matter}
\author{I. Brito$^\ast$, J. Carot$^\natural$, F. C. Mena$^\ast$, E.G.L.R. Vaz$^\ast$}

\maketitle
\address{$^\ast$Centro de Matem\'{a}tica, Departamento de Matem\'{a}tica e Aplica\c{c}\~{o}es,\\
Escola de Ci\^{e}ncias, Universidade do Minho, 4800-058 Guimar\~{a}es, Portugal\\
$^\natural$Departament de F\'{\i}sica, Universitat de les Illes
Balears,\\ Cra Valldemossa pk 7.5, E-07122 Palma de Mallorca, Spain
}

\begin{abstract}
The Einstein field equations are derived for a static cylindrically symmetric spacetime with elastic matter. The equations can be reduced to a system of two nonlinear ordinary differential equations and we present analytical and numerical solutions satisfying the dominant energy conditions. Furthermore, we show that the solutions can be matched at a finite radius to suitable $\Lambda$-vacuum exteriors given by the Linet-Tian spacetime.

\end{abstract}


\section{Introduction}
\label{Introduction}

Recently, there has been an increasing interest in the study of neutron stars in astrophysics using relativistic elasticity theory \cite{ChH},\cite{ZBH},\cite{PASHJ},\cite{GDSFM}. In fact, since neutron stars were found to be in elastic states \cite{McD}, \cite{Hae}, \cite{Pines}, there has been a number of developments and applications of elasticity theory in general relativity, mainly using spherically and axially symmetric elastic spacetimes to model astrophysical objects.
Results on spherically symmetric spacetimes in relativistic elasticity were
presented by Park \cite{Park} or by Magli and Kijowski \cite{MK}, \cite{M1}, showing that in spherical symmetry, the anisotropy in
pressures arises naturally as the relativistic extension of the
classical (non-relativistic) non-isotropic stress in elasticity
theory. Also Karlovini and Samuelsson \cite{KS1}, \cite{KS3} applied
the theory of elasticity to spherically symmetric spacetimes and
studied radial and axial perturbations \cite{KS2}, \cite{KS4}. Beig and Schmidt \cite{BS}
have shown that the Einstein field equations (EFEs) for elastic matter can be cast into a first-order
symmetric hyperbolic system and that local-in-time existence and uniqueness theorems
may be obtained under various circumstances. In
\cite{BCV} static and non-static shear-free solutions
have been obtained for the elastic EFEs.
However, one can find few results in the literature concerning exact solutions with
axially symmetric elastic configurations, due to the complexity of
the involved metrics. Magli \cite{M2} investigated the equilibrium
problem for axially symmetric, uniformly rotating neutron stars, while Calogero and Heinzle
\cite{Calogero} studied elastic spatially
homogeneous locally rotationally symmetric spacetimes in cylindrical symmetry.

In order to approach the problem of studying axially symmetric
elastic configurations we consider, as a first step, the case of
static cylindrical symmetry.


Using the formalism for relativistic elasticity proposed in \cite{MK}, \cite{M1},
the energy-momentum tensor is constructed for the static cylindrically symmetric elastic spacetime, depending on invariants of the pulled-back material metric. The EFEs are analysed, and following the technique indicated in \cite{BCV}, the EFEs are reduced to a system of two second-order ordinary differential equations. Particular examples of solutions are given, both analytically and numerically (for which the dominant energy conditions are satisfied). Furthermore, the matching problem of the interior elastic spacetime with the exterior Levi-Civita vacuum spacetime and its generalization including a cosmological constant, the Linet-Tian spacetime, is studied. It turns out that the matching across a cylindrical boundary is only possible with the latter spacetime.

The paper is organized as follows. Section 2 summarizes the formalism of relativistic elasticity. In Section 3, the static cylindrically symmetric elastic spacetime configuration is presented and the elastic energy-momentum tensor is written in terms of the eigenvalues of the pulled-back material metric. In Section 4, the EFEs are transformed into a system of two second-order ordinary differential equations. In Section 5, particular examples of solutions are presented and the dominant energy conditions are also analysed. In Section 6, the matching of the interior static cylindrically symmetric spacetime to the Levi-Civita exterior and to the Linet-Tian exterior is investigated.


\section{Relativistic elasticity in brief}

Consider a spacetime $(M,g)$, where $M$ is a 4-dimensional
Hausdorff, simply connected manifold of class $\mathcal{C}^2$ at least, and
$g$, a Lorentz metric of signature $(-,+,+,+)$. Coordinates in $M$ will be denoted as $x^a$ for $a=0,1,2,3$.
Let $X$ be a three-dimensional manifold representing the collection of particles of an elastic material, the \emph{material
space}, which is endowed with a Riemannian metric $\gamma$, the \emph{material metric}.
The material coordinates in $X$ will be denoted as $y^A$, $A=1,2,3$.

The spacetime configuration of the material is described by the submersion $\psi: M \rightarrow X$ through the three fields
$$  y^A = y^A(x^b), \qquad A=1,2,3.$$ The physical laws
describing the mechanical properties of the material can then be
expressed in terms of a hyperbolic second order system of PDEs. The
differential map $\psi_*: T_pM \rightarrow T_{\psi(p)} X $ can be
represented by the rank 3 matrix
$$ \left( y^A_{\; b}\right)_p, \qquad  y^A_{\; b}=\frac{\dep y^A}{\dep
x^b} \qquad A=1,2,3, \;\; b=0,1,2,3$$ which is called
\emph{relativistic deformation gradient}. Since $\psi_*$ has maximal
rank 3, its kernel is spanned at each point by a single timelike
vector which we may take as normalized to unity. The resulting
vector field, $\vec u = u^a\dep_a$, satisfies
\begin{equation}y^A_{\; b} u^b = 0, \quad u^au_a = -1, \;\; u^0 >0\label{uvm}\end{equation} the last
condition stating that it is future oriented. $\vec u$  is
called the \emph{velocity field of the matter}, and in the above picture in
which the points in $X$ are material points, it turns out that the
spacetime manifold $M$ is then made up by the worldlines of the
material particles, whose tangent vector is $\vec u$.

The pulled-back material metric given by $k_{ab}=(\psi^*
\gamma_{AB})=y^A_{\; a}y^B_{\; b}\gamma_{AB}$ and the related
operator $K^{a}_{\hspace{0.15cm}b}=-u^{a}u_{b}+k^{a}_{\hspace{0.15cm}b}$ can be used to measure the state of strain of the material, thus the material is said to be in a locally relaxed
state at $p\in M$ if the \emph{strain tensor} defined by $S_{ab} = -\frac12(k_{ab} - h_{ab})=-\frac12(k_{ab}-u_au_b
- g_{ab})$, where $h_{ab} = g_{ab} + u_au_b$, vanishes at $p$; or equivalently, if $K_{ab} = g_{ab}$ at $p$.

The strain tensor determines the elastic energy stored in an
infinitesimal volume element of the material space (or energy per
particle), hence that energy will be a scalar function of $K_{ab}$.
This function is called \emph{constitutive equation} of the
material, and its specification amounts to the specification of the
material. We shall represent it as $v = v(I_1,I_2,I_3)$, where
$I_1,I_2,I_3$ are any suitably chosen set of scalar
invariants associated with and
characterizing $K_{ab}$ completely. Notice that one of the eigenvalues of $K^a_{\hspace{0.15cm} b}$ is 1,
therefore, there exist three other scalars (in particular they could
be chosen as the remaining eigenvalues) characterizing $K^a_{\hspace{0.15cm} b}$
completely along with its eigenvectors. Following \cite{M1}, we shall
choose
\begin{equation}\label{I1.0} {I_{1}}=\frac{1}{2}\left(\text{Tr} {K}-4\right),\qquad
 {I_{2}}=\frac{1}{4}\left[\text{Tr} {K}^{2}-\left(\text{Tr} {K}\right)^{2}\right]+3,\qquad
 {I_{3}}=\frac{1}{2}\left(\text{det} {K}-1\right).\end{equation}
 The energy density $\rho$ will then be the particle number density
$\epsilon$ times the constitutive equation, that is \be \label{rho0}
\rho = \epsilon v(I_1,I_2,I_3) = \epsilon_0 \sqrt{\det K}\,
v(I_1,I_2,I_3),\ee where $\epsilon_0$ is the particle number density
as measured in the material space, or rather, with respect to the
volume form associated with $k_{ab} = (\psi^* \gamma)_{ab}$, and
$\epsilon$ is that with respect to $h_{ab}$; see \cite{KJ1} for a
proof of the above equation. In some references (e.g. \cite{M1}),
the names $\rho$ and $\epsilon$ are exchanged and the density
measured w.r.t. $k_{ab} = (\psi^* \gamma)_{ab}$ ($\epsilon_0$ in our
notation) is then called ``density of the relaxed material'' (see
the above comments on the meaning of $\gamma$), whereas that
measured w.r.t. $h_{ab}$ is referred to as the ``density in the rest
frame''.

The energy-momentum tensor for elastic matter is obtained from the Lagrangian $\mathcal{L} =
\sqrt{-g} \rho$, which depends on $y^{A}$, $y^{A}_{a}$ and $x^{a}$.
The corresponding Euler-Lagrange equations assume the form $\displaystyle{\frac{\partial\mathcal{L}}{\partial y^{A}}-\partial_{a}\left( \frac{\partial \mathcal{L}}{\partial y^{A}_{a}}\right)=0}$.
Using Noether's theorem one constructs the canonical energy-momentum tensor
\begin{equation}\label{cemt} \tilde{T}^{a}_{\hspace{0.15cm}b}=\frac{1}{\sqrt{-g}}\frac{\partial \mathcal{L}}{\partial y^{A}_{a}} y^{A}_{b}-\delta^{a}_{b}\mathcal{L},\end{equation}
which satisfies the conservation law $\nabla_{a}\tilde{T}^{ab}=0$.
The symmetric energy-momentum tensor $T_{ab}$ is the negative of the canonical energy-momentum tensor (see for instance \cite{KM94}). The energy-momentum tensor can be written in terms of the invariants of $K_{ab}$ as \cite{M2}
\begin{equation} \label{Tab}
T^{a}_{\hspace{0.15cm}b}=-\rho\,\delta^{a}_{b}+\frac{\partial\rho}{\partial I_{3}}\,
\text{det}K\,h^{a}_{\hspace{0.15cm}b}- \left(\text{Tr}
K\,\frac{\partial\rho}{\partial I_{2}}-\frac{\partial\rho}{\partial I_{1}}\right)k^{a}_{\hspace{0.15cm}b}+
\frac{\partial\rho}{\partial I_{2}}\,k^{a}_{\hspace{0.15cm}c}\,k^{c}_{\hspace{0.15cm}b}.
\end{equation}
This expression will be used to construct the EFEs for elastic matter $G^{a}_{\hspace{0.15cm}b}=T^{a}_{\hspace{0.15cm}b}.$

\section{Static cylindrically symmetric elastic configuration}

Consider a static cylindrically symmetric spacetime $(M,g)$, with
metric $g$ given by the line-element \be
ds^{2}=-e^{2\nu}dt^{2}+e^{2\mu}dr^{2}+e^{2\mu}dz^{2}+e^{2\psi}d\phi^{2},\label{csm}
\ee where the spacetime coordinates are $x^{a}=(t,r,z,\phi)$ and
$\nu$, $\mu$ and $\psi$ are $\mathcal{C}^2$ functions of $r$. The
associated material space $X$ is assumed to be such that the
configuration mapping $\psi: M \rightarrow X$ preserves the Killing
vectors (KVs), so that if $\vec{\xi_{A}}$ are KVs in $M$, where
$\vec{\xi_{1}}=\partial_{t}$, $\vec{\xi_{2}}=\partial_{z}$,
$\vec{\xi_{3}}=\partial_{\phi}$, then
$\psi_*(\vec{\xi_{A}})=\vec{\eta_{A}}$ are also KVs in $X$.
Therefore, the material metric $\gamma$ is also cylindrically
symmetric and it can be shown that  coordinates
$y^{A}=(R,\zeta,\Phi)$ exist in  $X$ with $R=R(r)$, $\zeta=z$ and
$\Phi=\phi,$ such that the material metric $\gamma$, assuming it to
be flat, can be represented by the line-element
$d\Sigma^{2}=dR^{2}+dz^{2}+R^{2}d\phi^{2},$ where $R=R(r)$. Here,
for simplicity, we restrict ourselves to the case in which $R(r)=r$,
so that $\gamma$ takes the form \be
\label{mm1}d\Sigma^{2}=dr^{2}+dz^{2}+r^{2}d\phi^{2}. \ee
If no such simplification is carried out, then, it is easy to see
(assuming flatness of the material metric $\gamma$) that the line
element \eqref{csm} above reads \be
ds^{2}=-e^{2\nu}dt^{2}+e^{2\mu}dR^{2}+e^{2\lambda}dz^{2}+e^{2\psi}d\phi^{2},\ee
where $\nu, \mu, \lambda$ and $\psi$ are functions of the radial
coordinate $R$. We shall not consider this case, though, and
restrict ourselves to the form \eqref{csm}.

In the case of the metric $\eqref{mm1}$ one
has:
\begin{equation}\begin{split}
k^{a}_{\hspace{0.15cm}b}&=g^{ac}k_{cb}=g^{ac}\left(y^{C}_{c}y^{B}_{b}\gamma_{CB}\right)\\
&=e^{-2\mu}\delta^{a}_{\hspace{0.15cm}1}\delta^{1}_{\hspace{0.15cm}b}+e^{-2\mu}\delta^{a}_{\hspace{0.15cm}2}
\delta^{2}_{\hspace{0.15cm}b}+r^{2}e^{-2\psi}\delta^{a}_{\hspace{0.15cm}3}\delta^{3}_{\hspace{0.15cm}b}.
\end{split}\end{equation}
The velocity field of the matter defined by \eqref{uvm} takes the form $u^{a}=\left(e^{-\nu(r)},0,0,0\right)$.
The operator $K^{a}_{\hspace{0.15cm}b}$ is given by
\begin{equation}K^{a}_{\hspace{0.15cm}b} =
\left(\begin{array}{cccc}1&0&0&0\\
0&e^{-2\mu}&0&0\\0&0&e^{-2\mu}&0\\0&0&0&r^{2}e^{-2\psi}\end{array}
\right), \label{K}\end{equation}
and we denote
\begin{equation}\label{etas} \eta =\,e^{-2\mu},\;\; \tau = \,r^{2}e^{-2\psi}.\end{equation}
The three invariants $I_1, I_2, I_3$ given in \eqref{I1.0}, written in terms of the eigenvalues of $K$, have the following expressions
\begin{equation}\label{I1}I_{1}=\frac{1}{2}\left(2\eta + \tau -3\right),\;\;
I_{2}=-\frac12 \left(\eta^{2} +2 \eta\tau +2 \eta +\tau\right) +3,\;\;
I_{3}=\frac12\left( \eta^{2}\tau -1\right).\end{equation}
For the given spacetime configuration, using these invariants, it is straightforward to show,  that the non-zero components of the energy-momentum tensor \eqref{Tab} are
\begin{eqnarray*}
T^{0}_{\hspace{0.15cm}0}&=&\,-\rho,\\
T^{1}_{\hspace{0.15cm}1}&=&\,-\rho+\frac{\partial\rho}{\partial I_{3}}\;\eta^{2} \tau -\frac{\partial\rho}{\partial I_{2}}(1+\eta+\tau)\eta+\frac{\partial\rho}{\partial I_{1}}\eta,\\
T^{2}_{\hspace{0.15cm}2}&=&T^{1}_{\hspace{0.15cm}1},\\
T^{3}_{\hspace{0.15cm}3}&=&\,-\rho+\frac{\partial\rho}{\partial I_{3}}\;\eta^{2}\tau-\frac{\partial\rho}{\partial I_{2}}(1+2\eta)\tau+\frac{\partial\rho}{\partial I_{1}}\tau,
\end{eqnarray*}
where the rest frame energy per unit volume $\rho$ is defined by \be
\rho=\epsilon v=\epsilon_{0}\,\sqrt{\text{det}K}v=\,\epsilon_{0}\,\sqrt{\eta^{2} \tau}v=\,\epsilon_{0}\,\eta \sqrt{ \tau}v.
\label{rfe}\ee
From \eqref{I1} it follows that
\begin{equation*}\begin{split}\label{drhoeta} &\frac{\partial\rho}{\partial\eta}=\frac{\partial\rho}{\partial I_{1}}-(1+\eta+\tau)\frac{\partial\rho}{\partial I_{2}}+\eta\tau \frac{\partial\rho}{\partial I_{3}}\\
&\frac{\partial\rho}{\partial\tau}=\frac{1}{2}\frac{\partial\rho}{\partial I_{1}}-\left(\eta+\frac{1}{2}\right)\frac{\partial\rho}{\partial I_{2}}+\frac{1}{2}\eta^{2}\frac{\partial\rho}{\partial I_{3}}.\end{split}\end{equation*}
Using these last results, the components of the energy-momentum tensor can be rewritten as:
\begin{equation}
\label{Tab2}
T^{0}_{\hspace{0.15cm}0}=-\epsilon v,\qquad
T^{1}_{\hspace{0.15cm}1}=T^{2}_{\hspace{0.15cm}2}=\epsilon \eta\frac{\partial v}{\partial \eta},\qquad
T^{3}_{\hspace{0.15cm}3}=2\epsilon \tau\,\frac{\partial v}{\partial \tau}.
\end{equation}

\section{The Einstein field equations}

Considering the results obtained in the previous section, the Einstein field equations $G^{a}_{\hspace{0.15cm}b}= T^{a}_{\hspace{0.15cm}b}$ for the above static cylindrically symmetric elastic configurations can be written as follows:
\begin{eqnarray}
G^{0}_{\hspace{0.15cm}0}= T^{0}_{\hspace{0.15cm}0}: \;&\frac{\mu''+\psi''+\psi'^{2}}{e^{2\mu}}&= \;-\epsilon v,\label{T00}\\
G^{1}_{\hspace{0.15cm}1}= T^{1}_{\hspace{0.15cm}1}:\;&\frac{\mu'\nu'+\mu'\psi'+\nu'\psi'}{e^{2\mu}}&=\;\epsilon \eta\frac{\partial v}{\partial \eta}, \label{T11}\\
G^{2}_{\hspace{0.15cm}2}= T^{2}_{\hspace{0.15cm}2}:\; &-\;\frac{\mu'\nu'+\mu'\psi'-\nu'^{2}-\nu''-\nu'\psi'-\psi''-\psi'^{2}}{e^{2\mu}}&= \;\epsilon \eta\frac{\partial v}{\partial \eta},\label{T22}\\
G^{3}_{\hspace{0.15cm}3}=T^{3}_{\hspace{0.15cm}3}:\;& \frac{\nu'^{2}+\nu''+\mu''}{e^{2\mu}} &= \;2\epsilon \tau\,\frac{\partial v}{\partial \tau}.\label{T33}
\end{eqnarray}

From \eqref{T11} and \eqref{T22} it follows that
\be 2\mu'\nu'+2\mu'\psi'-\nu'^{2}-\nu''-\psi''-\psi'^{2}=0.\label{T11equalT22}\ee
Now, dividing \eqref{T11} and \eqref{T33} through by \eqref{T00} and setting $E \equiv\ln \eta$ and $T \equiv \ln \tau$, one gets
\be\label{EFE4} \frac{\dep \ln
v}{\dep E}= - \frac{\mu'\nu'+\mu'\psi'+\nu'\psi'}{\mu''+\psi''+\psi'^{2}}, \qquad \frac{\dep \ln
v}{\dep T}= -\frac12\, \frac{\nu'^{2}+\nu''+\mu''}{\mu''+\psi''+\psi'^{2}}\;.\ee
Let $f\equiv \ln v,$ so that $f=f(E,T),$ $\frac{\partial f}{\partial E}=A(r)$ and $\frac{\partial f}{\partial T}=B(r).$
From the expressions for $\eta$ and $\tau$ one has that
\be E=-2\mu, \qquad T=2\ln r -2\psi.\ee
Therefore
\begin{equation}
A(r)=-\frac{1}{2\mu'}f'\;,\qquad
B(r)=\frac{1}{2\left(\frac{1}{r}-\psi'\right)}f'.
\label{AB}\end{equation}
In order for a constitutive equation $v=v(\eta,\tau)$, or equivalently $f(E,T) =\ln v(E,T)$, to exist, it must be that
\be\frac{\dep^2 f}{\dep T \dep E} = \frac{\dep^2 f}{\dep E \dep T}.\ee
Therefrom it follows \be \mu' A'=-\left(\frac{1}{r}-\psi'\right)B'.\ee
Now, from \eqref{AB}, one obtains \be \mu' A=-\left(\frac{1}{r}-\psi'\right)B \label{AB11}\ee
and using the former result leads to \be A=k_{0}B,\label{AB2}\ee where $k_{0}>0$.
Substituting \eqref{AB2} into \eqref{AB11} gives \be k_{0}\mu=\psi-\ln r,\label{eq1}\ee
where an additive constant has been set equal to zero without loss of generality.
Furthermore, \eqref{AB2} implies \be \mu'\nu'+\mu'\psi'+\nu'\psi'=\frac{k_{0}}{2}(\nu'^{2}+\nu''+\mu'').\label{eq2}\ee
Now, substituting $\psi=k_{0}\mu+\ln r$, obtained from \eqref{eq1}, in \eqref{eq2} and \eqref{T11equalT22} one gets the following form of the metric
\be
ds^{2}=-e^{2\nu}dt^{2}+e^{2\mu}(dr^{2}+dz^{2})+r^2e^{2k_{0}\mu}d\phi^{2},\label{csm1}
\ee
and the non-zero EFEs are:
\be k_{0}\mu'^{2}+(1+k_{0})\mu'\nu'+(\mu'+\nu')\frac{1}{r}-\frac{k_{0}}{2}(\nu''+\nu'^{2}+\mu'')=0\ee
\be k_{0}(2-k_{0})\mu'^{2}-k_{0}\mu''+\frac{2}{r}(1-k_{0})\mu'-(\nu''+\nu'^{2})+2\mu'\nu'=0.\ee
Setting $\mu'=M$ and $\nu'=N$ and subtracting in the last system the second equation from the first, one gets the following system of ordinary differential equations (ODEs)
\begin{eqnarray}
N'&=&\frac{k_{0}^{2}}{k_{0}-1}M^{2}-N^{2}+2\frac{k_{0}}{k_{0}-1}MN+
2\frac{k_{0}}{k_{0}-1}\frac{1}{r}M+\frac{2}{k_{0}-1}\frac{1}{r}N
\label{N}\\
M'&=&-\frac{k_{0}^{2}-2k_{0}+2}{k_{0}-1}M^{2}-\frac{2}{k_{0}(k_{0}-1)}MN-
\frac{2}{r}\frac{k_{0}^{2}-k_{0}+1}{k_{0}(k_{0}-1)}M-\frac{2}{r}\frac{1}{k_{0}(k_{0}-1)}N\label{M}\end{eqnarray}
where $k_{0}\neq 0$ and $k_{0}\neq1$. We will study separately the case $k_{0}=1$.

The dominant energy condition (DEC) holds if $\rho\geq 0$ and $|p_{A}|\leq \rho,$ $A=1,2,3,$ which implies
\begin{align}
\label{Dee}
\rho\equiv & \;Z (k_0+1)-\frac{2k_0 M}{r}-k_{0}^{2}M^{2}\geq 0,\nonumber\\
\rho\pm p_{1}\equiv &\; Z(k_0+1)\pm M^{2}k_0(1\mp k_0)\pm MN(1+k_0)\pm \frac{M}{r}\pm \frac{N}{r}-\frac{2k_0 M}{r}\geq 0,\nonumber\\
 \rho+p_{3}\equiv &\; Z k_0-\frac{2k_0M}{r}-k_{0}^{2}M^{2}+\frac{k_{0}^{2}M^{2}}{k_{0}-1}+\frac{2k_{0}MN}{k_{0}-1}+
\frac{2k_{0}M}{(k_{0}-1)r}+\frac{2N}{(k_{0}-1)r}\geq 0,\\
\rho-p_{3}\equiv
&\; Z(2+k_0)-\frac{2k_0 M}{r}-k_{0}^{2}M^{2}-\frac{k_{0}^{2}M^{2}}{k_{0}-1}-\frac{2k_{0}MN}{k_0 -1}-
\frac{2k_0 M}{(k_0 -1)r}-\frac{2N}{(k_0 -1)r}\geq
0,\nonumber
\end{align}
where $\displaystyle{Z=\frac{(k_{0}^2-2k_{0}+2)M^2}{k_0 -1}+\frac{2MN}{k_0 (k_0 -1)}+
\frac{2M(k_{0}^2-k_0+1)}{rk_0(k_0 -1)}+\frac{2N}{rk_0 (k_0 -1)}}.$

A careful analysis of the geometry around the axis (see
\cite{Carot}) implies that in order for the metric to be regular
at the axis, one must have $$ g_{\phi \phi} = r^2 g^{o}_{\phi
\phi}$$ where $g^{o}_{\phi \phi} \neq 0$ on the axis $r=0$
necessarily, that is: $\exp(2k_0\mu)$ cannot be zero on the axis.
Further, the so called \emph{elementary flatness condition} is also
implied:
 \begin{eqnarray*}
  \lim_{r\rightarrow 0} \frac{\nabla_c (\xi_a \xi^a) \nabla^c(\xi_a \xi^a)}{4\xi_a \xi^a}= 1,\end{eqnarray*}
 where $\vec{\xi}=(0,0,0,1)$ is the axial Killing vector, that is
 \be
 \lim_{r\rightarrow 0} e^{2(k_0-1)\mu}(1+ k_0 r \mu ')^2= 1.\label{regc}\ee
Considering $k_0=1,$ this condition is satisfied if
$\underset{r\rightarrow 0}{\lim}\; r \mu '\;=\;0.$

For $k_0 \neq 1$ it is easy to show, taking into account the above
considerations, that one must have that ${\displaystyle \lim_ {r\to
0} r \mu' (r) =C}$ with $C\in R$, and then the value of $\mu$ at the
origin must be such that \eqref{regc} is satisfied.

For  $k_0=1$, it is possible to integrate the EFEs and test
(\ref{Dee}) and (\ref{regc}). We will do this in the next section
and summarise the results so far as:

\begin{lem} For $k_0\ne 1$ there exist static, cylindrically symmetric solutions to the
Einstein Field Equations of the form (\ref{csm1}), which are regular
at the axis, satisfy the Dominant Energy Condition and represent
elastic matter, if there are sufficiently smooth real functions $M$
and $N$ which satisfy \eqref{N}-\eqref{regc}.
\end{lem}

For $k_0=1$, the solutions to the EFEs are given by (\ref{sol1}) and
contain subclasses which satisfy (\ref{Dee}) but not (\ref{regc}).

By standard results of the theory of non-linear ODEs it is known
that for systems as (\ref{N})-(\ref{M}), under certain conditions,
local solutions exist and are unique. Furthermore, the solutions are
continuously dependent on the initial data (say $M(r_\star)$ and
$N(r_\star)$) as well as on the parameters ($k_0$, in this case).
So, e.g. if $M(r_\star)>0$ and $N(r_\star)>0$ then, by continuity,
$M$ and $N$ will stay positive in a sufficiently small neighbourhood
of $r_\star$.

Not all solutions to (\ref{N})-(\ref{M}) will satisfy the
energy conditions. However, for any $\mathcal{C}^1$ functions $M$ and $N$ such that $M(r)>0$ and $N(r)>0$ one can show,
by inspection, that conditions (\ref{Dee}) are satisfied provided
$1<k_0<2.$ We summarise this as:
\begin{prop}
Consider $1<k_0<2$, $r_{\ast}\in]0,r_{0}[$, for some $r_0>0$, and
$M, N$ two $\mathcal{C}^1$ real functions. Given $M(r_{\ast})=\varepsilon >0$
and $N(r_{\ast})=\eta>0,$ then
\begin{eqnarray}
M'(r_{\ast})&=&\frac{-1}{k_0-1}\left[\eta\left(\frac{2}{k_0}\varepsilon+
\frac{2}{r_{\ast}}\frac{1}{k_0}\right)+(k_{0}^{2}-2k_0+2)\varepsilon^2+
\frac{2}{r_{\ast}}\frac{k_{0}^2-k_0+1}{k_0}\varepsilon\right]<0\\
N'(r_{\ast})&=&\frac{1}{k_0-1}\left[k_{0}^2\varepsilon^2+
\frac{2k_0}{r_{\ast}}\varepsilon+\left(2k_0
\varepsilon+\frac{2}{r_\star}\right)\eta\right]-\eta^2
\end{eqnarray}
and the (unique) solution to the ODE system (\ref{N})-(\ref{M}) is
such that $M(r)>0$ and $N(r)>0$, for
$r\in]r_{\ast}-\Delta,r_{\ast}+\Delta[$, with $\Delta>0$
sufficiently small, and satisfies the DEC
(\ref{Dee}).
\end{prop}
Similar results for other sign possibilities of
$M(r_\star)$ and $N(r_\star)$ are harder to prove and
we provide numerical examples, in the next section and in the appendix, for $M(r_*)>0,
N(r_*)<0$ and $M(r_*)<0, N(r_*)<0$ and $M(r_*)<0, N(r_*)>0$.
\section{Particular Solutions}\label{SsolEFE}

Explicit solutions to the EFEs in this case are difficult to find.
One can see that $M(r)=N(r)=0$ is a trivial solution of the ODE
system, which renders a flat metric.

In the case $k_0=2$, we obtain the explicit solution $\mu=-\ln(r)$,
$\nu=\ln(-1+rc)-\ln(r)$, where $c$ is a positive constant and
$r>1/c$. However, $\rho <0$ and thus this solution does not satisfy
the weak energy condition (nor the axis regularity condition), and
we can therefore rule it out on physical grounds.

Next, we shall analyse the more interesting $k_0=1$ case.
 In this case, the system leads to
 \begin{equation}
 \label{sol1}
 \displaystyle{N=\frac{-M^{2}r-2M}{2+2Mr}},~~~~ \displaystyle{M=\frac{3c\pm\sqrt{3c^{2}+6cr^{3}}}{(r^{3}-c)r}},
 \end{equation}
 so that \be\begin{split}&\mu=\int \frac{3c\pm\sqrt{3}\sqrt{c(c+2r^{3})}}{r(r^{3}-c)} dr+ c_{1},\\&\nu=\int -\frac{6r^{3}c\pm2c\sqrt{3}\sqrt{c(c+2r^{3})}\pm\sqrt{3}
\sqrt{c(c+2r^3)}r^{3}+3c^{2}}{r(r^{3}-c)(r^{3}+2c\pm\sqrt{3}\sqrt{c(c+2r^{3})})}dr +c_{2},\end{split}\ee where either 1) $c> 0 \wedge  r^3\neq c$ or 2) $c<-2r^3$ must hold and $c_{1},c_{2}$ are arbitrary constants.\\
Considering case 2), where $c<-2r^3$, one obtains $\rho\leq 0$. Therefore, this case can be ruled out as well. Considering now case 1), we find:
\begin{eqnarray}
\rho&=& \frac{1}{e^{2\mu}} \left[-\frac{1}{r^2}+\left(\frac{3c\pm\sqrt{3c^2+6cr^3}}{r(r^3 -c)}+\frac{1}{r}\right)^2\right] \nonumber\\
 p_1&=& \frac{1}{e^{2\mu}}\;\frac{(\sqrt{3}\sqrt{c(c+2r^3)} \pm3c)^2 }{2(r^3+2c\pm\sqrt{3}\sqrt{c(c+2r^3)}\;)r^2(-r^3 +c)}\\
p_3&=&\frac{1}{e^{2\mu}} \;\frac{6c\left[\sqrt{3}(4c^3+10c^2 r^3+4cr^6)   \pm \sqrt{c(c+2r^3)}(7c^2+10cr^3+r^6)\right]}{r^2 (-r^3+c)(r^3+2c\pm \sqrt{3}\sqrt{c(c+2r^3)}\;)^2\sqrt{c(c+2r^3)}}\nonumber
\end{eqnarray}
where $c> 0 \wedge r^3\neq c$. For $M$ with a negative sign, one
gets $\rho\leq 0.$ However, for $M$ with a positive sign, one can
show that the dominant energy condition is satisfied if and only
if $\displaystyle{r>\frac{1}{c^{1/3}}}.$
The regularity condition is not satisfied, since $\underset{r\rightarrow 0}{\lim}\; r \mu '\;=-3\mp\sqrt{3}\neq\;0.$

For other particular values of $k_0$ one can solve the system of equations \eqref{N} and \eqref{M} numerically using, for example,
the Fehlberg fourth-fifth order Runge-Kutta method by providing certain initial conditions and specifying the value of $k_0$.
As an example, in the appendix are listed some numerical results for $M(r)$ and $N(r)$ considering $k_0=1.3$ and establishing the following initial conditions: $M(0.1)=1$, $N(0.1)=1.$ Part of the interest of that example is that it satisfies the conditions of Proposition 1 for some range of $r$. In particular, one can also verify that the numerical solution in the appendix satisfies the DEC in the same ranges of $r$.
Numerical results for $k_0=0.5$ and $k_0=5$ are also given in the appendix.

\section{Matching to an exterior}
The aim of this section is to study the possibility of matching the cylindrically symmetric elastic spacetime \eqref{csm1} obtained in the previous section to a suitable exterior.
We shall consider two natural candidates for the exterior: the vacuum Levi-Civita spacetime \cite{LC} and its $\Lambda$-vacuum generalization, the so-called Linet-Tian spacetime \cite{L, T}.

The matching of any two $\mathcal{C}^2$ spacetimes $(M^\pm,g^\pm)$ with non-null boundaries $\Sigma^\pm$ requires the identification of the boundaries, i.e.  embeddings $\Phi_\pm: \Sigma\to M^\pm$ with $\Phi(\Sigma)=\Sigma^\pm$, where $\Sigma$ is an abstract copy of either boundary. We denote coordinates in $\Sigma$ by $\xi^\alpha$, $\alpha=1,2,3$, orthonormal tangent vectors to $\Sigma^\pm$ by $ e_\alpha^{\pm i}$ and normal vectors by $n_\pm^ i$. The first and
second fundamental forms at $\Sigma^\pm$ are $q^\pm_{\alpha \beta}=e_\alpha^{\pm i} e_\beta^{\pm j} g^\pm_{i j}$ and $H^\pm_{\alpha \beta}=-n^\pm_i e_\alpha^{\pm j} \nabla^\pm_ j e_\beta^{\pm i}$.
The necessary and sufficient conditions for the matching (in the absence of shells) are:
$$
q_{\alpha \beta}^-=q_{\alpha \beta}^+,~~~~H_{\alpha \beta}^-=H_{\alpha \beta}^+.
$$
A well know consequence of the matching conditions is the continuity of the normal components of the stress-energy tensor across the boundary, i.e. $n_i^- T^-_{i j}\eqq n_i^+ T^+_{i j}$, where $\eqq$ denotes equality at the boundary $\Sigma$. Taking as interior $(M^-,g^-)$ the elastic spacetime \eqref{csm1}
 of the previous section we must then expect, if the matching is possible, that $T^{1-}_1\eqq 0$ for a Levi-Civita exterior, and $T^{1-}_1\eqq -\Lambda$ for a Linet-Tian exterior.

Consider coordinates $\xi^a=\{ \lambda, \varphi, \xi \}$ on $\Sigma$ and vectors  tangent to $\Sigma^-$, adapted to the Killing vectors of $g^-$, given by
$$
e_1^-=\partial_t, ~~~e_2^-=\partial_z,~~~e_3^-=\partial_\phi.
$$
The matching will be performed across time-like surfaces which have constant radius, say $r_\Sigma$, and are geodesic. So we can set $\dot t=1$, where the overdot denotes a derivative with respect to $\lambda$. So, $\Sigma^{-}$ can be parameterized as $\Phi^-:=\{ t=\lambda, r=r_{\Sigma}, \phi=\varphi, z=\xi \}$. The first fundamental form at $\Sigma^-$ is then
$$
q^-=-e^{2\nu} d\lambda^2+e^{2\mu}d\xi^2 +e^{2\psi} d\varphi^2.
$$
In all cases below, the normal vector to $\Sigma^-$ will be taken to be
$$
n_-^i=e^{-\mu} \partial_r
$$
so that the only non-zero components of the second fundamental form at $\Sigma^-$ are:
$$
H^-_{\lambda\lambda}=-\nu' e^{2\nu-\mu}, ~~~~~H^-_{\xi\xi}=\mu' e^{\mu},~~~~~ H^-_{\varphi\varphi}=\psi' e^{2\psi-\mu}.
$$
\subsection{Impossibility of matching to a Levi-Civita exterior}
 The exterior Levi-Civita metric, $g^{+}$, is taken to be (see \cite{LC}) \be ds^{2}=-a^{2}\rho^{4\sigma}dT^{2}+b^{2}\rho^{4\sigma(2\sigma-1)}(d\rho^2+dz^2)+c^{2}\rho^{2(1-2\sigma)}d\phi^{2},\label{LeviC}\ee
where $a$, $b$, $c$ and $\sigma$ are real constants. Since the spacetime is static, we take the surface $\Sigma^+$ as $\rho=\rho_{\Sigma}$.
The tangent space to $\Sigma^+$ is then generated by the following basis vectors
$$ e_{1}^{+}=\partial_T,e_{2}^{+}=\partial_z,e_{3}^{+}=\partial_\phi$$
The unit normal vector to $\Sigma^+$ is
$$n^{i}_{+}=b^{-1}\rho^{-2\sigma(2\sigma-1)}\partial_{\rho}.$$
Calculating the first fundamental form on $\Sigma^+$ yields
$$q^+=-a^{2}\rho^{4\sigma}d\lambda^{2}+b^{2}\rho^{4\sigma(2\sigma-1)}d\xi^2
+c^{2}\rho^{2(1-2\sigma)}d\varphi^{2}$$
and the equality of the first fundamental forms implies
\be
e^{2\nu} \eqq a^{2}\rho^{4\sigma}\label{FF11}\ee
\be e^{2\mu} \eqq b^{2}\rho^{4\sigma(2\sigma-1)}\label{FF12}\ee
\be r^{2}e^{2k_{0}\mu} \eqq c^{2}\rho^{2(1-2\sigma)},\label{FF13}
\ee
wherefrom one gets
\be r^{2} \eqq c^{2}b^{-2k_{0}}\rho^{2(1-2\sigma)+4\sigma(2\sigma-1)k_{0}}.\label{FF14}\ee
Equating the second fundamental forms leads to
\be
-\nu'e^{-\mu+2\nu} \eqq
-\frac{2a^{2}\sigma}{b}\rho^{-4\sigma^{2}+6\sigma-1}\label{FF21}\ee
\be\mu'e^{\mu} \eqq 2b\sigma(2\sigma-1)
\rho^{4\sigma^2-2\sigma-1}\label{FF22}\ee
\be(r^{2}k_{0}\mu'+r)e^{2k_{0}\mu-\mu} \eqq
-\frac{c^{2}(2\sigma-1)}{b}\rho^{-4\sigma^2-2\sigma+1}.\label{FF23}
\ee
The first and second fundamental forms enable to write $\mu_{\Sigma},\nu_{\Sigma},\mu'_{\Sigma},\nu'_{\Sigma},r_{\Sigma}$ and $k_{0}$ as functions of quantities of the exterior spacetime.

Using \eqref{FF12}, \eqref{FF14} and \eqref{FF22} one obtains from \eqref{FF23} the following formula for $r_{\Sigma}$:
\be r \eqq \frac{\rho}{(2\sigma-1)(2k_{0}\sigma+1)}.\label{FF5}\ee
From \eqref{FF12}, \eqref{FF22} and \eqref{FF23}, one gets
\be k_{0}=-\frac{1}{2\sigma}.\label{ksig}\ee
Condition \eqref{ksig} implies from \eqref{T11} a zero radial pressure at the boundary $\Sigma$. However, as it can be seen, \eqref{ksig} and \eqref{FF5} are incompatible.
Thus, we conclude:
\begin{lem}
It is impossible to match the cylindrically symmetric elastic static
spacetime \eqref{csm1} to a Levi-Civita exterior \eqref{LeviC}
across a timelike cylindrical surface of constant radius.
\end{lem}

The here obtained result showing the impossibility of matching an elastic fluid cylinder with the exterior Levi-Civita solution does not apply to perfect fluid cylinders.\\
In \cite{BLSZ}, Bi$\check{c}$\'{a}k et al. studied global properties of static perfect fluid cylinders and demonstrated the existence and uniqueness of global solutions for a general equation of state. There it is shown that the Levi-Civita solution can be joined to a static cylindrical perfect fluid interior if the pressure vanishes at a finite value of the radial coordinate.\\
The difference is that in the present case, when the fluid is elastic, which is more general than a perfect fluid because its energy-momentum tensor contains an anisotropic pressure tensor (see e.g. \cite{KS1}), the radial pressure does not vanish at the surface. In order for the radial pressure to vanish, the condition (\ref{ksig}) must hold, but as we showed in this section, this condition does not allow the possibility of matching, since it is incompatible with the matching conditions.

\subsection{The matching to a Linet-Tian exterior}
We now investigate the matching problem considering as exterior spacetime the $\Lambda$-vacuum Linet-Tian solution, for the EFEs $G_{ab}^{+}=-\Lambda g_{ab}^{+},$ in static cylindrical symmetry.

The Linet-Tian metric $g^{+}$ can be expressed as (\cite{L,T}, see also \cite{GP})
\be\label{Linet-Tian}\begin{split} ds^{2}_+=&-A^2 Q^{2/3}P^{-2(1-8\sigma+4\sigma^2)/3\Omega}dT^{2}+d\rho^2+\\
&B^2 Q^{2/3}(P^{-2(1+4\sigma-8\sigma^2)/3\Omega}dz^2+
C^{2}P^{4(1-2\sigma-2\sigma^2)/3\Omega}d\phi^{2}),\end{split}\ee
where $\Omega=1-2\sigma+4\sigma^2$ and, in the hyperbolic case,
\be \begin{split}
&Q(\rho)=\frac{1}{\sqrt{3|\Lambda|}}\sinh(\sqrt{3|\Lambda|}\rho)\\
&P(\rho)=\frac{2}{\sqrt{3|\Lambda|}}\tanh \left(\frac{\sqrt{3|\Lambda|}}{2}\rho\right),
\end{split}\ee
where $\Lambda< 0$ is the cosmological constant.
The solution has in general five free parameters, namely the cosmological constant $\Lambda$ and the real constant parameters $A,B,C\ne 0$ and $\sigma\ge 0$. The parameter $A$ can be interpreted as a time rescaling, $B$ and $C$ are related to the conicity and $\sigma$ is interpreted as the mass per unit length (see \cite{Silva-etal, GP}). The case $\Lambda>0$ is obtained by replacing the hyperbolic functions by trigonometric ones (see \cite{L},\cite{T}) and, for that case, the calculations in this section remain also valid.

We note that, although The Linet-Tian solution with $\Lambda<0$ is well behaved, for $\Lambda>0$ it has, besides the singularity at $\rho=0$ where we placed the source, another singularity at $\rho=\pi/\sqrt{3\Lambda}$ where another source has to be placed. In that case, the matching is still possible by substituting the cylindrical region by a toroidal one following the methods of \cite{GP}.


Sources for these solutions were studied in  \cite{Zofka-Bicak} as cylindrical shells of counter-rotating photons or dust and of perfect fluids (for different signs of $\Lambda$). A matching with no shells, and for $\Lambda>0$, was studied in \cite{GP} for an interior given by the Einstein static universe.

Here we shall match an exterior Linet-Tian solution to the static elastic solution of section \eqref{SsolEFE}. We start by taking a cylindrical surface $\Sigma^+$ in the exterior with tangent vectors
$$
e_1^+=\partial_T, ~~~e_2^+=\partial_z,~~~e_3^+=\partial_\phi
$$
such that $\Sigma^+$ is parameterized as $
\Phi^+:=\{ T=T(\lambda), \rho=\rho_\Sigma, \phi=\varphi, z=\xi \}
$
for a constant radius $\rho_\Sigma$. The matching is performed along a timelike geodesic for which we again set $\dot T=1$. The first fundamental form at $\Sigma^+$ is then
\be
q^+=  Q^{\frac{2}{3}}(-A^2P^{-2a} d\lambda^2+B^2P^{-2b}d\xi^2 +C^2 P^{2d} d\varphi^2)
\ee
with
$a=(1-8\sigma+4\sigma^2)/3\Omega,
 b=(1+4\sigma-8\sigma^2)/3\Omega,
  d=2(1-2\sigma-2\sigma^2)/3\Omega.$ Note that the following identities are satisfied
\begin{eqnarray}
\label{abd}a+b-d&\equiv&0\\
\label{abd2}a^2+b^2+d^2&\equiv&\frac{2}{3}.
\end{eqnarray}
The normal from the exterior  will be taken to be
$$
n_+^i=\partial_\rho
$$
so that the only non-zero components of the second fundamental form at $\Sigma^+$ are:
\begin{eqnarray}
H^+_{\lambda\lambda}&=&A^2 P^{-2a} Q^{\frac{2}{3}}\left(-\frac{1}{3}\frac{Q'}{Q}+a\frac{P'}{P}\right) \nonumber\\
H^+_{\xi\xi}&=& B^2 P^{-2b} Q^{\frac{2}{3}}\left(\frac{1}{3}\frac{Q'}{Q}-b\frac{P'}{P}\right) \nonumber\\
H^+_{\varphi\varphi}&=&C^2 P^{2d} Q^{\frac{2}{3}}\left(\frac{1}{3}\frac{Q'}{Q}+d\frac{P'}{P}\right) \nonumber
\end{eqnarray}
We use the following strategy: we take an interior solution $(M^{-},g^{-})$ which, by the results of Section \eqref{SsolEFE}, we know exists and is unique, given appropriate initial data.
This solution provides interior data $(\mu_\Sigma, \nu_\Sigma, \mu'_\Sigma, \nu'_\Sigma)$ at some boundary hypersurface $\Sigma$ where $r=r_\Sigma$.  The matching conditions will then fix, at the boundary,  the exterior data $\Lambda, A, B, C, \rho_\Sigma $ and $\sigma$ in terms of that interior data.

The equality of the first fundamental forms imply
\begin{eqnarray}
\label{match1}
AQ^{\frac{1}{3}}P^{-a}&\eqq& e^{\nu}  \nonumber\\
 BQ^{\frac{1}{3}}P^{-b}&\eqq&e^{\mu}  \\
 CQ^{\frac{1}{3}}P^{d}&\eqq& e^{\psi} \nonumber
\end{eqnarray}
In turn, the equality of the second fundamental form gives,
using \eqref{match1} and \eqref{eq1},
\begin{eqnarray}
\label{match21}
\mu' e^{-\mu}&\eqq& \frac{1}{3} \frac{Q'}{Q}- b \frac{P'}{P}\nonumber\\
\nu' e^{-\mu}&\eqq& \frac{1}{3} \frac{Q'}{Q}- a\frac{P'}{P} \\
 \left(\frac{1}{r}+k_0 \mu'\right) e^{-\mu}&\eqq& \frac{1}{3} \frac{Q'}{Q}+d \frac{P'}{P}.\nonumber
\end{eqnarray}
and, after a
long but straight forward calculation, using the interior EFEs, give
\begin{equation}
\label{T11L}
T^{1-}_{~1}\equiv\frac{1}{e^{2\mu}}(\mu'\nu'+\mu'\psi'+\nu'\psi')\eqq
-\Lambda,
\end{equation}
as expected\footnote{Note that if the
interior radial pressure is negative, i.e. $T^{1-}_{~1}<0$, at the
boundary, then $\Lambda>0$. This is the case of some $k_0=1$ solutions
of Section \ref{SsolEFE}, case 1).
}. Note that since
$\Lambda\neq 0$, then $\mu'\nu'+\mu'\psi'+\nu'\psi'\neqq 0$, which
excludes the case $k_0=0$. In any other case, the interior data
$k_0, r_\Sigma, \mu'_\Sigma, \nu'_\Sigma$ and $\mu_\Sigma$ determine
$\Lambda\neq0$.

Since \eqref{abd} holds, it follows from (\ref{match21}) that
\begin{eqnarray}
\label{rho1}
\coth(\sqrt{3|\Lambda|}\rho)\eqq\frac{e^{-\mu}}{\sqrt{3|\Lambda|}}\left(\frac{1}{r}+\nu'+\mu'+k_0\mu'\right),
\end{eqnarray}
so, after $\Lambda$, the quantity $\rho_\Sigma$ is also determined by interior quantities only.
Now,  from the second and third equations in \eqref{match21}, one gets
 \begin{eqnarray}\label{sig}
\frac{1-4\sigma}{1-2\sigma+4\sigma^2}\frac{\sqrt{3|\Lambda|}}{\sinh(\sqrt{3|\Lambda|}\rho)}\eqq e^{-\mu}\left(\frac{1}{r}-\nu'+k_0\mu'\right)
\end{eqnarray}
and allows to determine $\sigma$ as
\begin{equation}
\label{sigma}
\sigma=\frac{1}{2}\left(\frac{1}{2}-\frac{\alpha}{\beta}\pm \sqrt{\left(\frac{1}{2}-\frac{\alpha}{\beta}\right)^2+\frac{\alpha}{\beta}-1} \right)
\end{equation}
for $\alpha=\sqrt{3|\Lambda|}/\sinh(\sqrt{3|\Lambda|}\rho_\Sigma)>0$ and $\beta=e^{-\mu_\Sigma}\left(\frac{1}{r_\Sigma}-\nu'_\Sigma+k_0\mu'_\Sigma\right)$. It is easy to see that there are positive real solutions for $\sigma$ if and only if $\left(\frac{\alpha}{\beta}\right)^2>3/4$.
The actual values of $\sigma$ can be determined once the interior data at $\Sigma$ is known.
Then, by substituting \eqref{T11L}, \eqref{rho1}, \eqref{sig} in (\ref{match1}),
the parameters $A$, $B$ and $C$ are also completely
determined by interior quantities only. This then fixes the exterior
spacetime, globally.

In particular, it can be shown that the numerical solutions presented in the appendix satisfy the matching conditions for some values of the parameters. Considering for example the case $k_0=1.3$ and taking $r_{\Sigma}=0.5$, see Table \ref{table1} and Table \ref{table2}, one obtains $\Lambda=0.4524$, $\rho_{\Sigma}=0.4024$, $\sigma=0.1121$ (the other value for $\sigma$ is negative), $A=2.7473$, $B=0.5351$ and $C=0.6506$ and the second matching conditions \eqref{match21} are identically satisfied. Other examples for e.g. a negative $\Lambda$ can easily be obtained.

We conclude that,
given an interior solution with data $(k_0, \mu_\Sigma, \nu_\Sigma, \mu'_\Sigma, \nu'_\Sigma)$, with $k_0\neq0$, at some $r=r_\Sigma$, the exterior parameters
 $\Lambda, A, B, C, \sigma$ and $\rho_\Sigma$ are determined from the matching conditions.
In this sense, the matching of the static elastic spacetime to a
Linet-Tian spacetime exterior, across a cylindrical surface, exists
and, given condition \eqref{sigma}, is not unique. We summarize the results of this
section as:
\begin{thm}
Consider an interior cylindrically symmetric static elastic
spacetime \eqref{csm1} with data $(k_0, \mu_\Sigma, \nu_\Sigma,
\mu'_\Sigma, \nu'_\Sigma)$, with $k_0\neq0$, at some hypersurface $\Sigma$ such that $r=r_\Sigma$.
The matching of this interior spacetime to an exterior cylindrically
symmetric static $\Lambda$-vacuum spacetime \eqref{Linet-Tian},
across $\Sigma$, exists and the exterior metric parameters
 $\Lambda, \rho_\Sigma, \sigma$ and $A, B, C$ are determined from the interior data at $r_\Sigma$ through conditions \eqref{T11L},  \eqref{rho1}, \eqref{sigma} and \eqref{match1}, respectively.
\end{thm}
The matched solution represents an elastic static cylindrical object
imbedded in a $\Lambda$-vacuum static cylindrical spacetime which is
asymptotically anti de-Sitter, if $\Lambda<0$, or asymptotically
de-Sitter, if $\Lambda>0$. The sign of $\Lambda$ is fixed by the
sign of the interior radial pressure at the boundary $\Sigma$.
Interestingly, our elastic solutions are the first known examples of non-conformally flat interiors to the Linet-Tian spacetime.
\\\\
{\bf Acknowledgments:}
The authors thank CMAT, Universidade do Minho, for support through the FEDER Funds - "Programa Operacional Factores de Competitividade  COMPETE" and by
Portuguese Funds through FCT - "Funda\c{c}\~{a}o para a Ci\^{e}ncia e a Tecnologia",
within the Project Est-C/MAT/UI0013/2011. FM is supported by projects PTDC/MAT/108921/2008 and CERN/FP/116377/2010. JC is grateful for the support of the European Union FEDER funds, the Spanish Ministry  of Economy and Competitiveness  (projects FPA2010-16495 and CSD2009-00064) and the Conselleria d'Economia Hisenda i Innovacio of the Govern de les Illes Balears.

\section{Appendix}
In this appendix we present numerical solutions of the system
\eqref{N} and \eqref{M} for $k_0=1.3$, $k_0=0.5$ and $k_0=5$.
The cases $k_0=0.5$ and $k_0=5$ provide examples for the cases $0<k_0<1$ and $k_0>2$ which were not contemplated in Proposition 1.
\\\\

{\bf Case $k_0=1.3$}

\begin{figure}[b]
    \includegraphics[width=9cm]{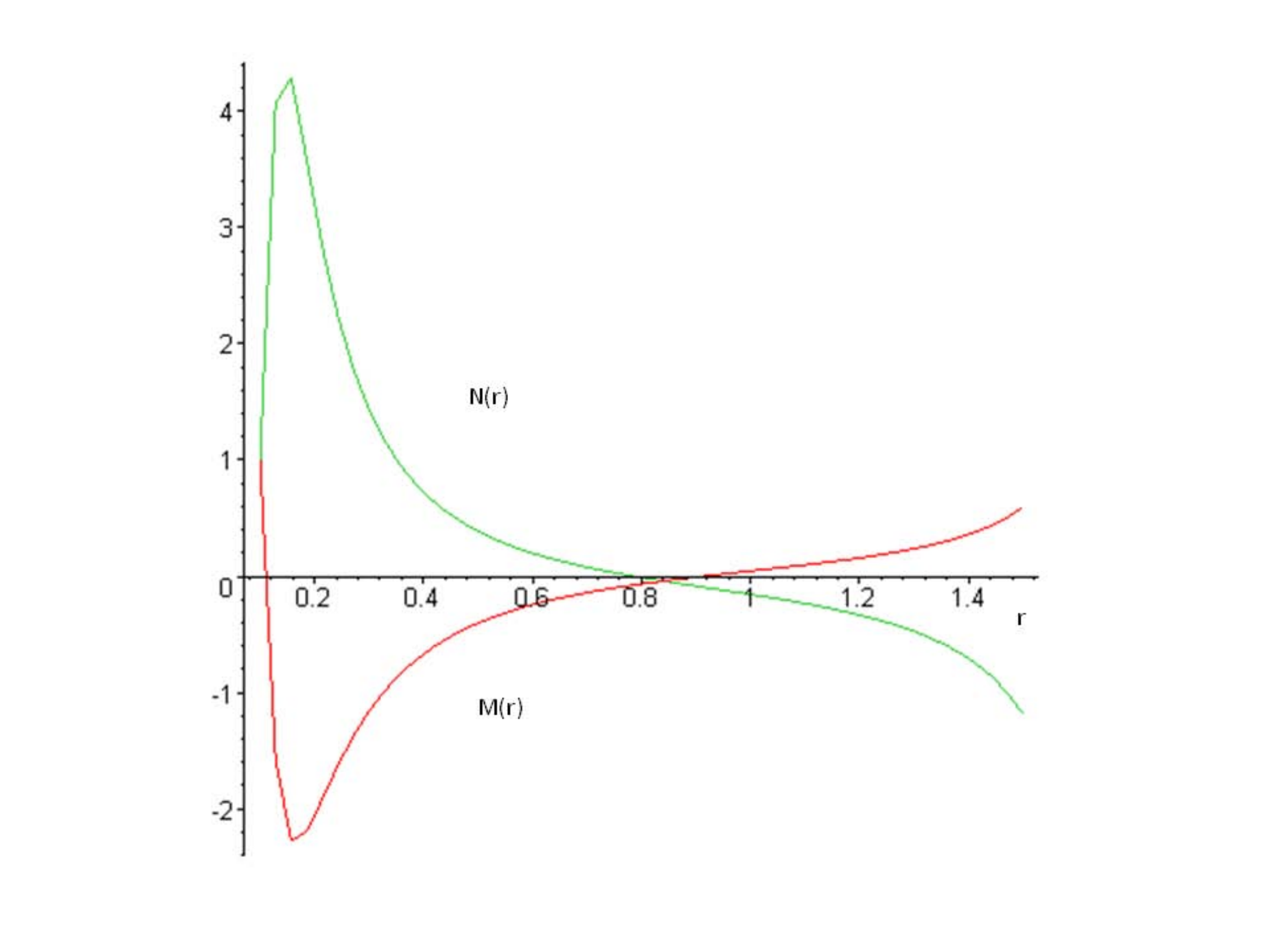}\\
  \caption{$M(r)$ and $N(r)$ for $k_0=1.3$ with ICs $M(0.1)=1$, $N(0.1)=1.$}\label{graphic1}
\end{figure}

In Table \ref{table1} are given some numerical results for $M(r)$ and $N(r)$ establishing the following initial conditions (ICs): $M(0.1)=1$, $N(0.1)=1.$ Figure \ref{graphic1} shows the numerical presentation of these results for $r\in[0.1,1.5]$.
In this case, one can verify that the numerical solution satisfies the DEC for certain ranges of $r$: $r\in(0,R),$ with $0.1<R<0.2.$

\begin{table}
\caption{Some numerical results for $k_0=1.3$ with ICs $M(0.1)=1$, $N(0.1)=1.$}
\begin{tabular}{|c|c|c|}
\hline $r$ & $M(r)$ & $N(r)$\\
\hline
0.102 & 0.743 & 1.326\\
0.105 & 0.379 & 1.789\\
0.108 & 0.407 & 2.215\\
0.2 & -2.062 & 3.193\\
0.3 & -1.162 & 1.433\\
0.4 & -0.669 & 0.722\\
0.5 & -0.403 & 0.385\\
1 & 0.045 & -0.157\\
1.5 & 0.596 & -1.187\\
\hline
\end{tabular}
\label{table1}
\end{table}

\begin{figure}
    \includegraphics[width=9cm]{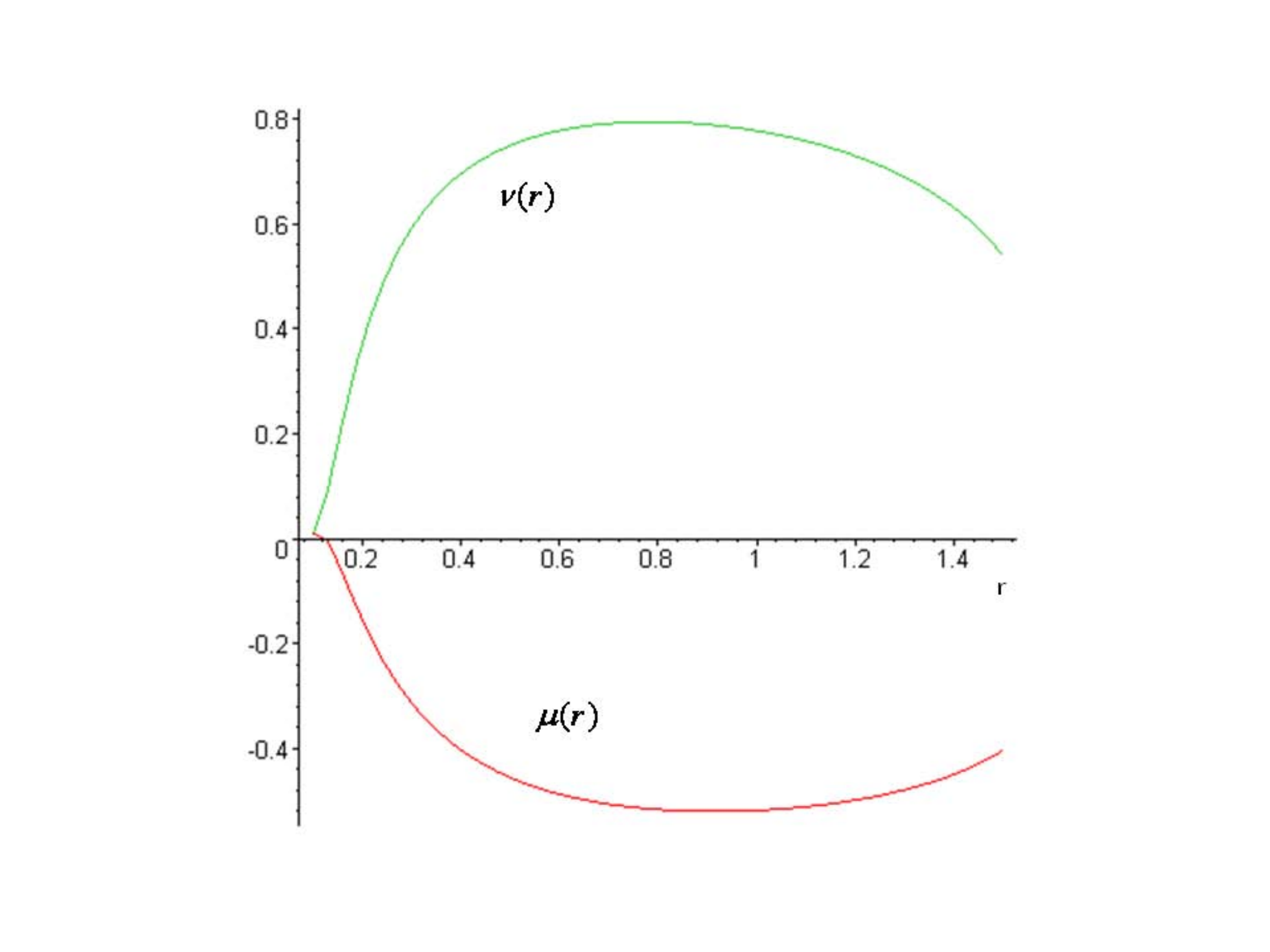}\\
  \caption{$\mu(r)$ and $\nu(r)$ for $k_0=1.3$ with ICs $\mu(0.1)=0.01$, $\nu(0.1)=0.01$, $\mu'(0.1)=1$, $\nu'(0.1)=1.$}\label{graphic2}
\end{figure}

In Table \ref{table2} are listed the results for $\mu(r)$ and $\nu(r)$ considering the ICs: $\mu(0.1)=0.01$, $\nu(0.1)=0.01$, $\mu'(0.1)=1$, $\nu'(0.1)=1$. Figure \ref{graphic2} contains the graphical presentation of $\mu(r)$ and $\nu(r)$ for values of $r\in[0.1,1.5].$

\begin{table}
\caption{Some numerical results for $k_0=1.3$ with ICs $\mu(0.1)=0.01$, $\nu(0.1)=0.01$, $\mu'(0.1)=1$, $\nu'(0.1)=1.$}
\begin{tabular}{|c|c|c|}
\hline $r$ & $\mu(r)$ & $\nu(r)$\\
\hline
0.102 & 0.012 & 0.012\\
0.105 & 0.013 & 0.017\\
0.108 & 0.014 & 0.023\\
0.2 & -0.157 & 0.376\\
0.3 & -0.315 & 0.594\\
0.4 & -0.404 & 0.697\\
0.5 & -0.456 & 0.750\\
1 & -0.519 & 0.778\\
1.5 & -0.404 & 0.541\\
\hline
\end{tabular}
\label{table2}
\end{table}

{\bf Case $k_0=0.5$}

\begin{figure}
    \includegraphics[width=9cm]{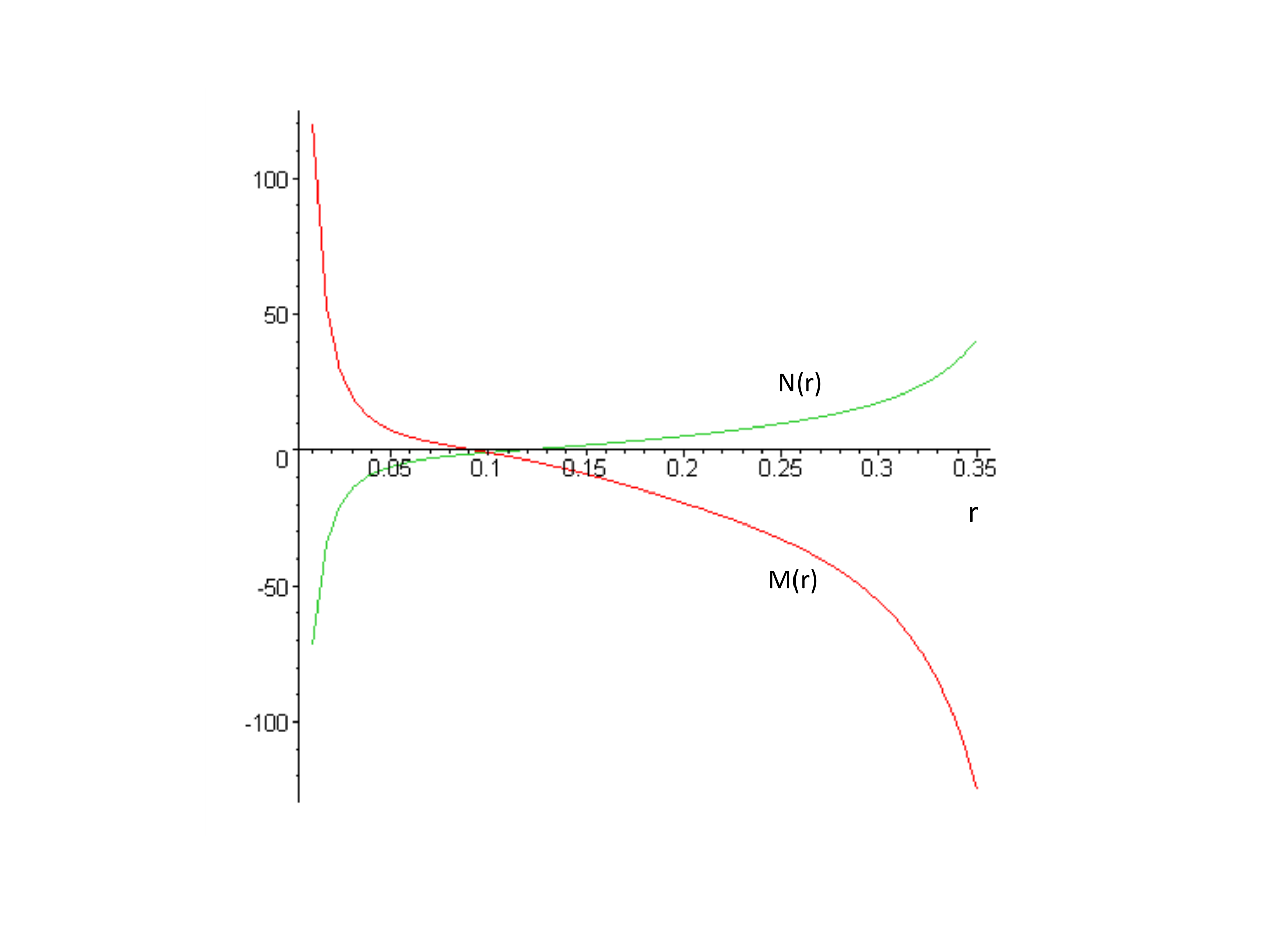}\\
  \caption{$M(r)$ and $N(r)$ for $k_0=0.5$ with ICs $M(0.1)=-1$, $N(0.1)=-1.$} \label{graphic3}
\end{figure}

Figure \ref{graphic3} shows the numerical presentation of the results for$M(r)$ and $N(r)$ considering $k_0=0.5$, $r\in[0.01,0.35]$ and establishing the following ICs: $M(0.1)=-1$, $N(0.1)=-1$.
One can verify that this numerical solution satisfies the DEC for $r\in(0.01,0.35).$

\begin{figure}
    \includegraphics[width=9cm]{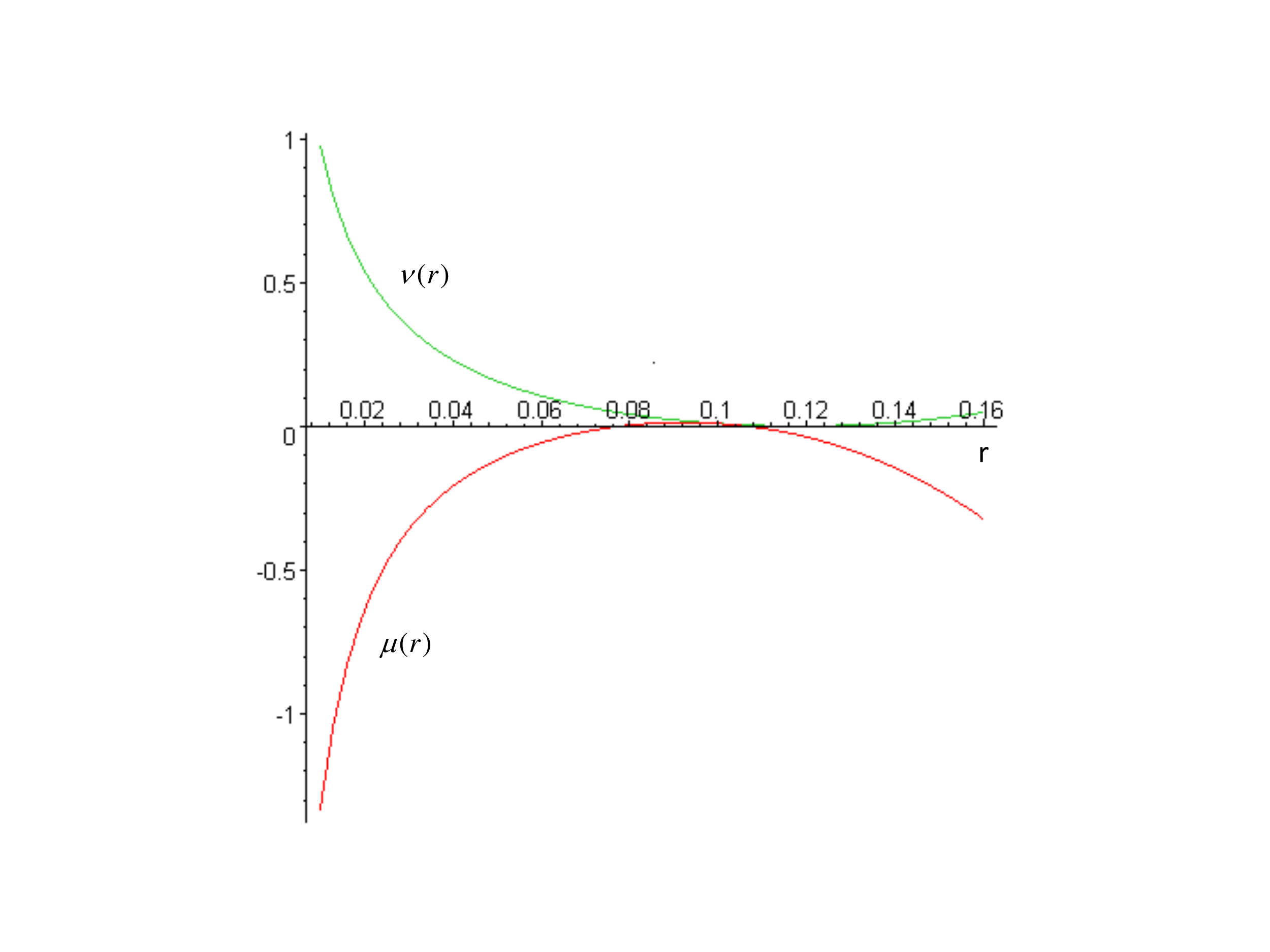}\\
  \caption{$\mu(r)$ and $\nu(r)$ for $k_0=0.5$ with ICs $\mu(0.1)=0.01$, $\nu(0.1)=0.01$, $\mu'(0.1)=-1$, $\nu'(0.1)=-1.$}\label{graphic4}
\end{figure}

Figure \ref{graphic4} contains the graphical presentation of $\mu(r)$ and $\nu(r)$ for values of $r\in[0.01,0.16],$ considering the ICs: $\mu(0.1)=0.01$, $\nu(0.1)=0.01$, $\mu'(0.1)=-1$, $\nu'(0.1)=-1$.

{\bf Case $k_0=5$}

\begin{figure}
    \includegraphics[width=9cm]{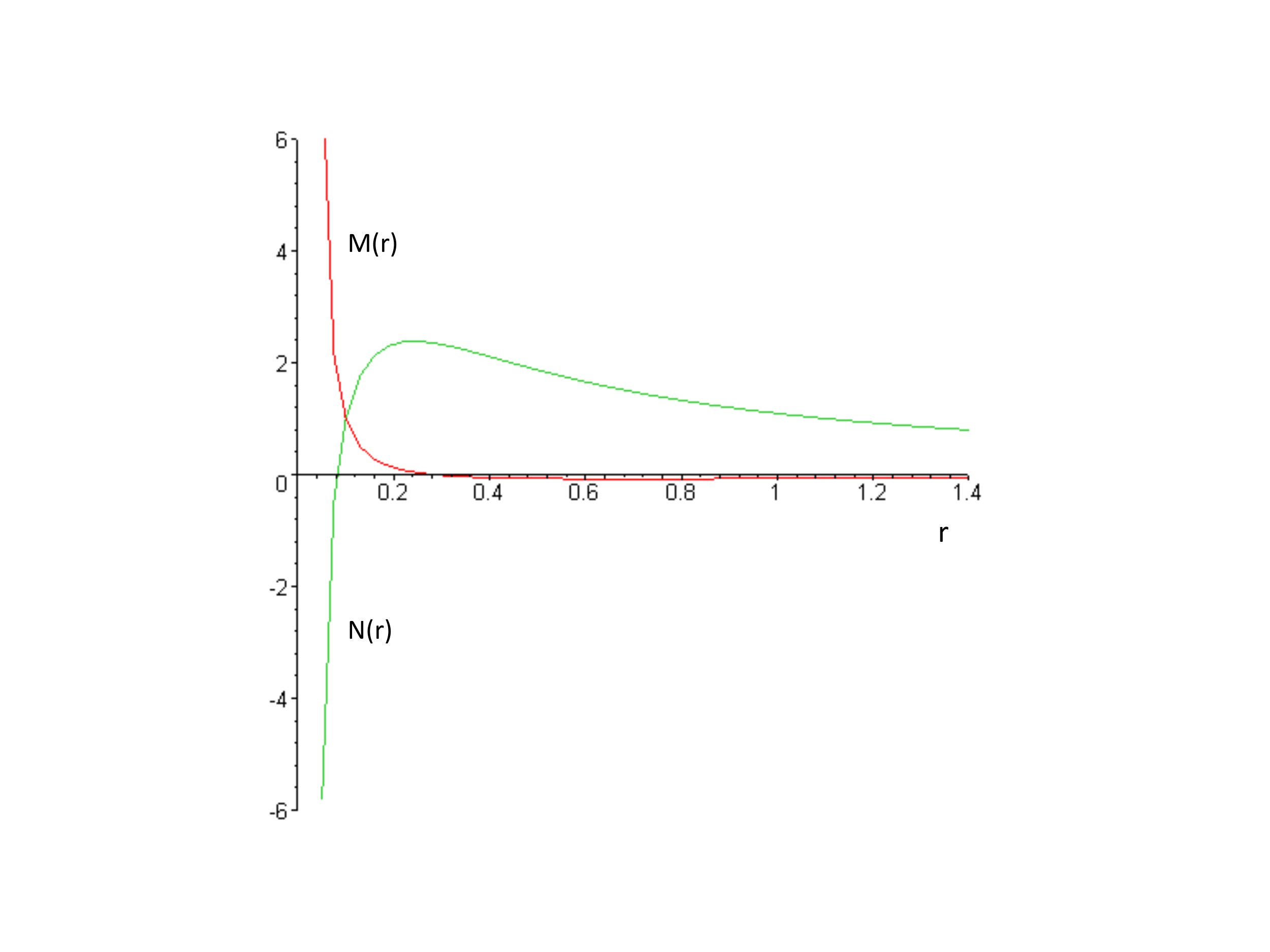}\\
  \caption{$M(r)$ and $N(r)$ for $k_0=5$ with ICs $M(0.1)=1$, $N(0.1)=1.$}\label{graphic5}
\end{figure}

Figure \ref{graphic5} shows the numerical presentation of the results for $M(r)$ and $N(r)$ considering $k_0=5,$ $r\in[0.04,1.4]$ and establishing the following ICs: $M(0.1)=1$, $N(0.1)=1.$  $r\in[0.04,1.4]$. One can verify that this numerical solution satisfies the DEC for $r\in(0.04,0.1).$

\begin{figure}
    \includegraphics[width=9cm]{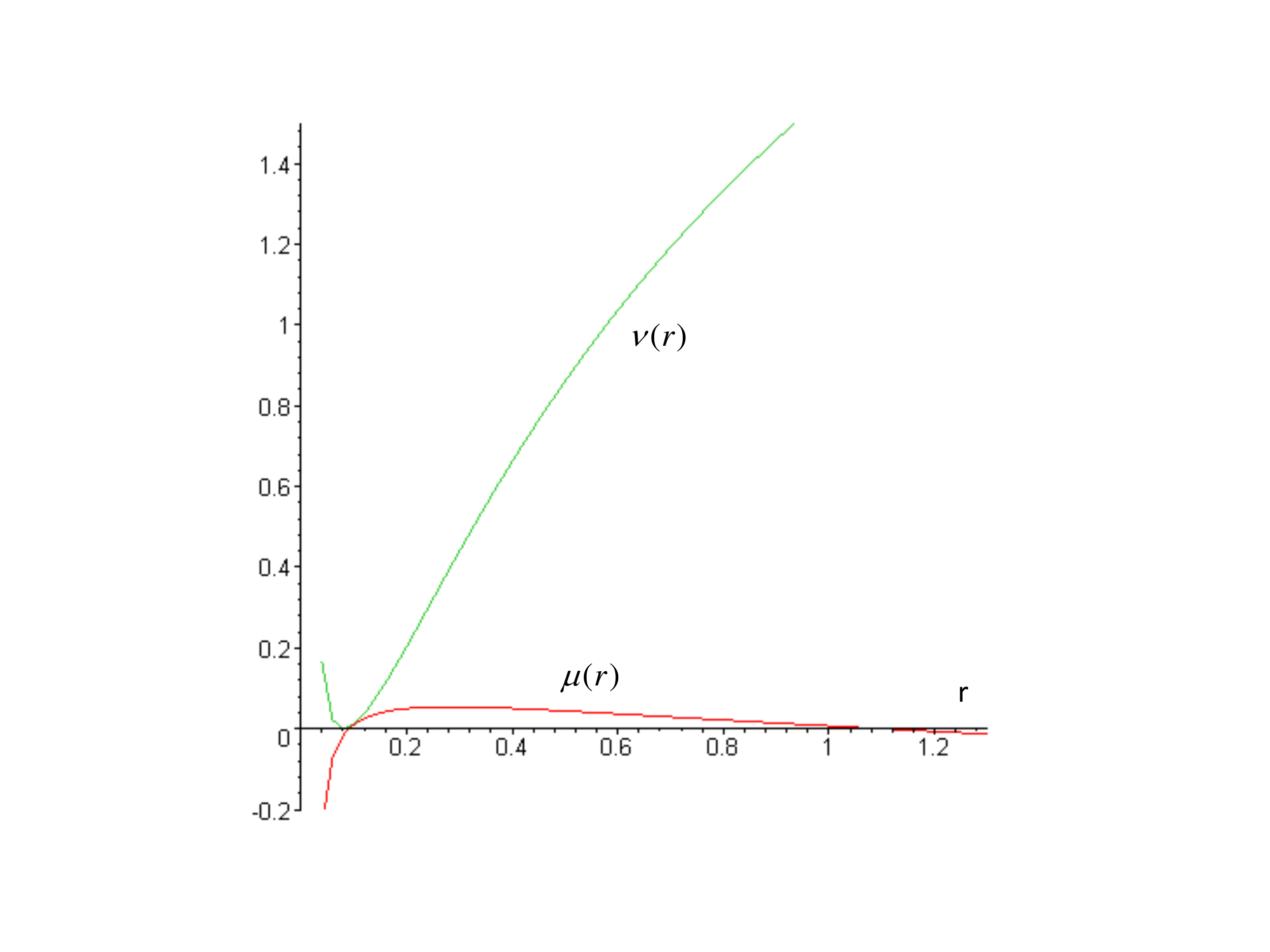}\\
  \caption{$\mu(r)$ and $\nu(r)$ for $k_0=5$ with ICs $\mu(0.1)=0.01$, $\nu(0.1)=0.01$, $\mu'(0.1)=1$, $\nu'(0.1)=1.$}\label{graphic6}
\end{figure}

Figure \ref{graphic6} contains the graphical presentation of $\mu(r)$ and $\nu(r)$ for values of $r\in[0.04,1.2]$ considering the ICs: $\mu(0.1)=0.01$, $\nu(0.1)=0.01$, $\mu'(0.1)=1$, $\nu'(0.1)=1$.

\end{document}